\documentclass[10pt]{iopart}

\RequirePackage[pdftex]{graphicx}
\usepackage{nomencl}
 \makenomenclature
 \usepackage{setspace}

\begin{document}

\title[Numerical analysis of azimuthal rotating spokes in a crossed-field discharge plasma]{Numerical analysis of azimuthal rotating spokes in a crossed-field discharge plasma}

\author{R Kawashima$^1$, K Hara$^2$ and K Komurasaki$^1$}

\address{$^1$The University of Tokyo, 7-3-1 Hongo, Bunkyo-ku, Tokyo 113-8656, Japan}
\address{$^2$Texas A\&M University, Bright Building, College Station, TX 77843, USA}
\ead{kawashima@al.t.u-tokyo.ac.jp}
\vspace{10pt}
\begin{indented}
\item[]November 2017
\end{indented}

\begin{abstract}
	Low-frequency rotating spokes are obtained in a cross-field discharge plasma using two-dimensional numerical simulations.
	A particle-fluid hybrid model is used to model the plasma flow in a configuration similar to a Hall thruster.
	It has been reported that the drift-diffusion approximation for an electron fluid results in an ill-conditioned matrix when solving for the potential because of the differences in the electron mobilities across the magnetic field and in the direction of the E$\times$B drift.
	In this paper, we employ a hyperbolic approach that enables stable calculation, namely, better iterative convergence of the electron fluid model.
	Our simulation results show a coherent rotating structure propagating in the E$\times$B direction with a phase velocity of 2,500 m/s, which agrees with experimental data.
	The phase velocity obtained from the numerical simulations shows good agreement with that predicted by the dispersion relation of the gradient drift instability.
\end{abstract}

%
%
\submitto{\PSST}
%
%
\ioptwocol
%


	\nomenclature[1l]{$L_z$}{axial channel length}
	\nomenclature[1l]{$L_\theta$}{azimuthal channel length}
	\nomenclature[1n]{$n$}{number density}
	\nomenclature[1b]{$B$}{magnetic flux density}
	\nomenclature[1t]{$T$}{temperature}
	\nomenclature[1u]{$u$}{velocity}
	\nomenclature[1k]{$k$}{wave number}
	\nomenclature[2$\phi$]{$\phi$}{space potential}
	\nomenclature[1u]{$u$}{velocity}
	\nomenclature[1m]{$m$}{mass}
	\nomenclature[1f]{$f$}{reaction rate coefficient}
	\nomenclature[1vp]{$v_{\rm p}$}{phase velocity}
	\nomenclature[1q]{$q_{\rm ion}$}{ion production rate}
	\nomenclature[2$\mu$]{$\mu$}{electron mobility}
	\nomenclature[2$\nu$e]{$\nu_{\rm ela}$}{elastic collision frequency}
	\nomenclature[2$\nu$i]{$\nu_{\rm ion}$}{ionization collision frequency}
	\nomenclature[2$\lambda$]{$\lambda$}{wavelength}
	\nomenclature[2$\varepsilon$i]{$\varepsilon_{\rm ion}$}{ionization energy}
	\nomenclature[2$\omega$]{$\Omega_{\rm e}$}{electron Hall parameter}
	\nomenclature[2$\tau$]{$\tau$}{pseudo time}
	\nomenclature[2$\omega$]{$\omega$}{angular frequency}
	\nomenclature[2$\lambda_{\rm D}$]{$\lambda_{\rm D}$}{Debye length}
	\nomenclature[2$\omega_{\rm p}$]{$\omega_{\rm p}$}{plasma frequency}
	\nomenclature[b0]{Subscript}{}
	\nomenclature[be]{e}{electron}
	\nomenclature[bi]{i}{ion}
	\nomenclature[bn]{n}{neutral}

\section{Introduction}

	Rotating spokes are often observed in a cross-field configuration, where the electric field and magnetic field are perpendicular to each other.
	Examples of cross-field devices include magnetron discharges \cite{BoeufPRL2013,Brenning:2013aa} and Hall effect thrusters (HETs) \cite{EllisonPoP2012,SekerakTPS2015}. 
	Azimuthally rotating spokes are self-organizing structures that exhibit long-wavelength and low-frequency plasma oscillations on the order of 10--100 kHz.
	Such rotating structures typically propagate in the E$\times$B direction and its propagation speed is significantly  (say, two or three orders of magnitude) slower than the electron E$\times$B drift. 
	Depending on the plasma properties and geometry of the electromagnetic fields, the rotating direction can be either along ($+$) or opposite ($-$) of the E$\times$B drift \cite{Hecimovic:2016aa,ItoTPS2008}.
	
	
	It has been reported that the discharge conditions in Penning and magnetron discharges affect the characteristics of the spokes, such as the number of modes and propagation speed, similar to HETs \cite{Hecimovic:2016aa,Ito:2015aa}.
	While researchers have proposed several hypotheses for these source, including critical ionization velocity (CIV) model \cite{Brenning:2013aa} and Simon-Hoh instability \cite{Sakawa:1993aa}, it is still unclear what the source of the coherent structure is.
	Part of the reason for the poor understanding is because of the lack of simulations of low-frequency plasma oscillations in the E$\times$B direction.
	While a high-fidelity kinetic simulation is required for the detailed calculations of rotating spokes, a fluid model, particularly for electrons, is required for a long-time large-scale simulation that captures the low-frequency dynamics.
	
	One of the main challenges in understanding the physics of cross-field discharge plasmas is the large disparity in time scales from ps to ms.
	For a HET, the plasma frequency is on the order of a few tens of GHz while the rotating spoke frequency is a few kHz.
	Therefore, it is computationally challenging to employ non-neutral conditions, i.e., resolving the plasma frequency.
	In modeling low-frequency phenomena, a fluid model is often used for electrons assuming a quasi-neutral plasma.
	Such models have been used to investigate breathing mode, another low-frequency plasma oscillation resulting in discharge current oscillation \cite{Boeuf:1998aa,Gascon:2003aa,Barral:2003aa}.
	It was suggested in Ref. \cite{Hara:2014aa,Hara:2014ab} that the electron dynamics (e.g., heat balance and convection) play an important role in such ionization instabilities.	
	For rotating spoke simulations, Lam et al. and Fernandez et al. have attempted axial-azimuthal simulations using a hybrid model assuming quasineutrality \cite{Lam6922570,Fernandez2015iepc}.
	The azimuthal oscillations shown were not as clear as the ones observed in experiment and the authors reported that the simulations were not stable after a given period.
	These results infer that there is a numerical instability in axial-azimuthal coordinate systems.
	Hara and Boyd performed a 2D hybrid model using a direct kinetic model for heavy species to eliminate the numerical noise inherent in particle methods \cite{HaraIEPC2015}.
	It was found that the heavy species modeling (numerical noise) does not play an important role, but the potential equation obtained from the  electron current conservation using the drift-diffusion approximation is ill-conditioned in the axial-azimuthal calculation.
	This is because the electron propagation speed becomes several orders of magnitude different between the cross-field (axial) and E$\times$B (azimuthal) directions when electrons are strongly magnetized.
	
	In this paper, an axial-azimuthal hybrid numerical simulation is developed to model the azimuthally rotating, long-wavelength and low-frequency structure in a cross-field discharge plasma.
	A hyperbolic system approach is used for the electron fluid calculation to avoid the numerical instability that arises when solving an elliptic partial differential equation \cite{Hagelaar:2007aa,Kawashima201559}.
	The method used in this paper is a robust computational method for magnetized electrons that converts a diffusion equation (elliptic partial differential equation) into a hyperbolic system by using a pseudo-time advancement technique \cite{Kawashima201559,Kawashima2016202,Kawashima:2016aa}.
	In Sections 2 and 3, the axial-azimuthal plasma model and numerical method are described. 
	Results of the rotating spokes are discussed in Section 4.
	In Section 5, the mechanism of rotating spokes is investigated by applying linear theories to the numerical results, and the growth rate and phase velocity are assessed.
	Finally, we propose a multidimensional spoke model based on the numerical results.

\section{Axial-azimuthal particle-fluid hybrid model}
\begin{figure}[t]
		\begin{center}
			\includegraphics[width=80mm]{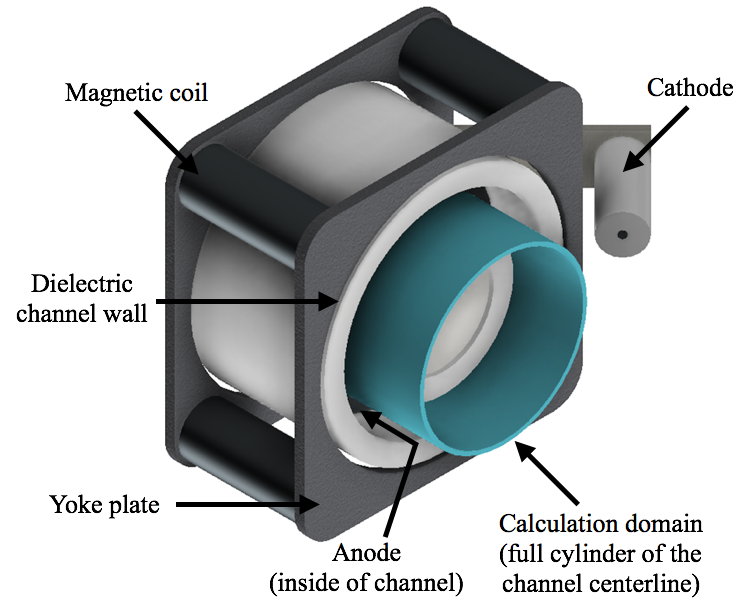}
		\end{center}
		\caption{Configuration of an HET and the calculation domain assumed in the simulation.}
		\label{fig:domain}
	\end{figure}

	The configuration of a HET is shown in Fig. \ref{fig:domain}.
	A calculation domain of the full azimuth in an HET is assumed for modeling the discharge plasma between the anode and cathode in the axial direction.
	Further, to investigate the low-frequency rotating spokes, a long-time simulation of several milli-seconds is performed.
	Long-time simulations using a non-neutral fully kinetic model, such as a particle-in-cell (PIC) model, is infeasible given the required grid resolution and time step.
	The ratio of spoke wavelength to Debye length becomes $\lambda_{\rm spoke}/\lambda_{\rm D}\sim 10^4$ and the ratio of spoke frequency to plasma frequency is $\omega_{\rm spoke}/\omega_{\rm plasma}\sim 10^6$.
	Hence, a large number of grid points and temporal iterations will be required to fully capture the low-frequency azimuthally rotating spokes using a fully kinetic simulation.
	
	To achieve a long-time calculation with a reasonable computational cost, a quasineutral particle-fluid hybrid model is adopted.
	Ions and neutral atoms are modeled as particles, whereas the electrons are modeled as a fluid assuming a quasineutral plasma.
	In numerical simulations of HETs, such hybrid models have been one of the most popular approaches owing to the fidelity of modeling and reasonable computational cost \cite{FifeThesis,Komurasaki:1995fk}.
	The basics of the hybrid model is described below.

\subsection{Ion and neutral atoms}
	The ion and neutral atoms are modeled using a PIC method.
	The effects of ion magnetization is neglected for simplicity.
	Thus, the ions are accelerated only electrostatically.
	The ionization and electron-neutral elastic collision frequencies are calculated by the following equations:
	\begin{equation}
		\nu_{\rm ion} = n_{\rm n}f_{\rm ion,Xe}\left(T_{\rm e}\right),\hspace{15pt}
		\nu_{\rm ela} = n_{\rm n}f_{\rm ela,Xe}\left(T_{\rm e}\right).
		\label{eq:collision}
	\end{equation}
	where $\nu_{\rm ion}$, $\nu_{\rm ela}$, $n_{\rm n}$, and $T_{\rm e}$ are the ionization collision frequency, electron-neutral elastic collision frequency, neutral atom density, and electron temperature.
	$f_{\rm ion,Xe}$ and $f_{\rm ela,Xe}$ are the ionization and elastic reaction rate coefficients for xenon, respectively.
	These reaction rate coefficients are given as approximate functions of electron temperature assuming a Maxwellian electron energy distribution \cite{GoebelKatz2008}.
	The neutral atoms are provided from the anode and the inflow neutral atom flux is determined by the anode mass flow rate.
	Inter-particle collisions are neglected, and all the neutral atoms move to the downstream without collisions.
	Ions are generated in the domain via ionization.
	Ions colliding with the anode neutralize and reflect back into the domain diffusively as neutral atoms.
	In this axial-azimuthal model, the physics in the radial direction, i.e., plasma-wall interaction, is neglected.

\subsection{Electron fluid}
	
	An electron fluid model consists of conservations of mass, momentum, and energy.
	First, the electron mass conservation is given in a steady-state form
	\begin{equation}
		\nabla\cdot\left(n_{\rm e}\vec{u}_{\rm e}\right)=q_{\rm ion},
		\label{eq:mass}
	\end{equation}
	where $n_{\rm e}$, $\vec{u}_{\rm e}$, and $q_{\rm ion}$ are the electron density, electron velocity, and ionization rate, respectively.
	Here, $q_{\rm ion}=n_{\rm e}\nu_{\rm ion}$.
	Since electrons are assumed to be highly mobile to instantaneously achieve the quasineutrality, the time-derivative term of electron number density is excluded from the mass conservation.
	The validity of the quasineutral assumption can be considered by comparing the grid sizing with the Debye length.
	If the Debye length $\lambda_{\rm D}$ is much smaller than the grid sizing $\Delta z$, the charge separation are considered to be shielded within each cell.
	In the present simulation, the electron number density and electron temperature are on the order of 10$^{18}$ m$^{-3}$ and 10 eV, respectively.
	Thus, the Debye length is on the order of $\lambda_{\rm D}\sim10^{-5}$ m.
	On the other hand, the grid sizing of the present simulation is $\Delta z\sim10^{-3}$ m.
	Hence $\lambda_{\rm D}\ll \Delta z$, and the quasineutral assumption is considered to be valid in the electron fluid model.
	For the analyses of microscopic phenomena in the scale smaller than $\lambda_{\rm D}$, the quasineutral assumption may not be valid and the time-derivative term of electron number density could be included in the mass conservation.
	
	For the electron momentum conservation, a drift-diffusion approximation (generalized Ohm's law) is used:
	\begin{equation}
		-n_{\rm e}\left[\mu \right]\nabla \phi+\left[\mu\right]\nabla\left(n_{\rm e}T_{\rm e}\right)=-n_{\rm e}\vec{u}_{\rm e},
		\label{eq:momentum}
	\end{equation}
	where $\phi$ is the space potential.
	$\left[\mu\right]$ is the electron mobility tensor that accounts for magnetization.
	In the axial-azimuthal coordinates,
	\begin{equation}
	   \left[\mu\right]
		=\mu_\perp\left[
		\begin{array}{cccc}
		1 & \Omega_{\rm e} \\
		-\Omega_{\rm e}  & 1
		\end{array}
		\right],
		\label{eq:mu_zt}
	\end{equation}
	where $\mu_\perp$ and $\Omega_{\rm e}$ are the electron mobility across the magnetic field lines and electron Hall parameter, respectively, which are explained in the next section in detail.
	From Eqs. (\ref{eq:mass}) and (\ref{eq:momentum}), an equation for plasma potential can be written as follows:
	\begin{equation}
		\nabla\cdot\left(n_{\rm e}\left[\mu\right]\nabla\phi\right)=
		\nabla\cdot\left(\left[\mu\right]\nabla\left(n_{\rm e}T_{\rm e}\right)\right)
		+q_{\rm ion},
		\label{eq:elliptic}
	\end{equation}
	which is a second-order elliptic partial differential equation for the potential.
	
	The electron energy conservation is calculated in the axial direction to derive an axial electron temperature distribution.
	Assuming that the electron temperature is uniform in the azimuthal direction, two-dimensional plasma properties of electron density, axial electron flux, electron mobility, space potential, and ionization rate are averaged in the azimuthal direction, and the obtained axial plasma profile is used in the energy conservation equation.
	The 1D electron energy conservation in the axial direction is formulated as follows:
	\begin{eqnarray}
		\frac{\partial}{\partial t}\left(\frac{3}{2}e\bar{n}_{\rm e}\bar{T}_{\rm e}\right)
		+\frac{\partial}{\partial z}\left(\frac{5}{2}e\bar{\Gamma}_{{\rm e},z}\bar{T}_{\rm e}
		-\frac{5}{2}e\bar{n}_{\rm e}\bar{T}_{\rm e}\bar{\mu}_{\rm \perp}\frac{\partial \bar{T}_{\rm e}}{\partial z}\right) \nonumber\\
		=e\bar{\Gamma}_{{\rm e},z}\frac{\partial \bar{\phi}}{\partial z}
		-e\alpha_{\rm E} \varepsilon_{\rm ion}\bar{q}_{\rm ion}.
		\label{eq:energy}
	\end{eqnarray}
	where $e$, $\Gamma_{\rm e}$, and $\varepsilon_{\rm ion}$ are the elementary charge, electron number flux, and ionization energy for xenon, respectively. $\alpha_{\rm E}$ is a dimensionless coefficient accounting for the energy losses due to inelastic collisions, and it is defined as $\alpha_{\rm E}=2.0+0.254\exp\left(2\varepsilon_{\rm ion}/3T_{\rm e}\right)$. \cite{Dugan_Sovie_1967}.
	The quantities with bars in Eq. (\ref{eq:energy}) mean that they are averaged in the azimuthal direction.	
	A similar 1D electron temperature solver was used in the axial-azimuthal hybrid PIC simulation by Lam et al. \cite{Lam6922570}.
	Although the variation of $T_{\rm e}$ in the azimuthal direction was observed experimentally \cite{EllisonPoP2012}, the effect of $T_{\rm e}$ fluctuation in the azimuthal direction on a rotating spoke is considered to be small according to the linear perturbation analysis \cite{FriasPoP2013}.
	Further, it is confirmed from our numerical simulations that the axial electron temperature profile is almost in a steady state, and breathing-type axial ionization oscillation is not produced.

\subsection{Cross-field electron mobility}
	In the HET discharge plasma, it is known that the electron mobility based on the classical diffusion is not sufficient to obtain the discharge current and plasma properties observed in experiments \cite{Hofer:2008aa}.
	For this, anomalous electron mobility is often introduced in the cross-field direction.
	Recent studies suggest that high-frequency plasma waves, such as the electron cyclotron drift instability and ion acoustic waves, may be the source of cross-field electron transport \cite{Lafleur:2016ab,Boeuf:2017aa}.
	
	In this study, an empirical electron transport model is used for the cross-field transport for two reasons.
	1) As mentioned previously, the anomalous transport is required to establish plasma profiled that are similar to the HET discharge plasma observed in experiments.
	2) Axial ionization oscillations, namely, the breathing mode, can be damped in the presence of anomalous electron transport \cite{HaraThesis}.
	The latter is particularly important for our study because the presence of breathing mode can complicate the azimuthally rotating spokes.
	As shown later, spokes occur in the absence of breathing mode oscillations, indicating that azimuthally rotating spokes are not a combination of nor driven by axial ionization oscillations.
	Note that the anomalous transport model typically only describes the cross-field transport, and its effects to the parallel or E$\times$B direction are often not discussed.
	Hence, we model the electron mobilities as follows.
	The cross-field electron mobility is given by,
	\begin{equation}
		\mu_{\perp}=\mu_{\rm \perp,cla}+\mu_{\rm \perp,ano},
		\label{eq:muperp}
	\end{equation}
	where $\mu_{\rm \perp,cla}$ and $\mu_{\rm \perp,ano}$ are the classical and anomalous electron mobilities, respectively.
	The classical electron mobility is written as
	\begin{equation}
		\mu_{\rm \perp,cla}=\frac{\mu_{||}}{1+\Omega_{\rm e,cla}^2}\sim\frac{\mu_{||}}{\Omega_{\rm e,cla}^2},
		\label{eq:mu_cla}
	\end{equation}
	where $\mu_{||}$ is the mobility along the magnetic field lines and $\Omega_{\rm e,cla}=\mu_{||}B$ is the electron Hall parameter based on collisions.
	In this model, $\mu_{||}$ is defined as non-magnetized electron as follows:
	\begin{equation}
		\mu_{||}=\frac{e}{m_{\rm e}\nu_{\rm ela}},
	\end{equation}
	where $m_{\rm e}$ is the electron mass.
	The anomalous electron mobility is given by a Bohm-type diffusion model as follows:
	\begin{equation}
		\mu_{\rm \perp,ano}=\frac{\alpha_{\rm B}}{16B},
	\end{equation}
	where $\alpha_{\rm B}$ is an empirical coefficient. 
	In HET modelings, numerous models have been proposed for the distribution of the $\alpha_{\rm B}$ \cite{Hofer:2008aa,Hagelaar:2002aa,KooPoP2006}.
	In the present model, the distribution of the $\alpha_{\rm B}$ is determined by using the Gaussian function as follows:
	\begin{equation}
		\alpha_{\rm B} =-C\exp\left(-\frac{1}{2\sigma^2}\left(\frac{z}{z_{\rm c}}-1\right)^2\right)+D,
	\end{equation} 
	where $z_{\rm c}$ is the channel length. 
	The coefficients $C$, $D$, and $\sigma$ are assumed as follows:
	\begin{equation}
		\left\{\begin{array}{c}
			C = 0.12,\ D = 0.14,\ \sigma = 0.0050,\hspace{10pt}
			 \left(\frac{z}{z_{\rm c}} \leq 1\right)\\
			C = 0.98,\ D = 1.00,\ \sigma = 0.0075.\hspace{10pt}
			 \left(\frac{z}{z_{\rm c}} > 1 \right)
		\end{array}\right.
	\end{equation}
	These coefficients are chosen based on the three-region model proposed in Ref. [26].
	In the original three-region model, there exist indifferentiable points in the distribution of $\alpha_{\rm B}$.
	Indifferentiable points in $\alpha_{\rm B}$ may numerically excite artificially instabilities.
	Hence, the Gaussian function in Eq. (11) is introduced in the present model to avoid indifferentiable points.
	The present cross-field electron mobility model was examined by using an axial-1D hybrid simulation in advance of the 2D simulation, and a good agreement was confirmed between the cross-field electron mobility obtained form the axial-1D simulation and the empirical data \cite{Lafleur:2016ab}.
	
	The effects of anomalous transport on the electron mobility in the E$\times$B direction are not well understood.
	In this paper, we assume that Eq. (\ref{eq:mu_zt}) and Eq. (\ref{eq:mu_cla}) hold in the presence of anomalous transport,
	and the electron Hall parameter including the effects of anomalous transport is defined as $\Omega_{\rm e} = \sqrt{\mu_{\rm ||}/\mu_{\rm \perp}}$.

\subsection{Hyperbolic equation system approach}

	The hyperbolic-equation system (HES) approach is a robust approach to solving elliptic partial differential equations by converting them into a hyperbolic equation system.
	The off-diagonal elements can be much greater than the diagonal elements in Eq. (\ref{eq:mu_zt}) since $\Omega_{\rm e}\gg 1$.
	It was shown in Ref. \cite{Kawashima201559} that the existence of the off-diagonal elements degrades the diagonal dominance in the coefficient matrix, and computation becomes unstable. 
	The details of the HES and numerical procedures are discussed in Ref. \cite{Kawashima201559}.
	
	In the HES approach, a hyperbolic system of electron conservation laws is constructed by introducing pseudo-time advancement terms.
	For instance, the hyperbolic system for the electron mass and momentum conservations in Eqs. (\ref{eq:mass}) and (\ref{eq:momentum}) can be given by
	\begin{equation}
		-\frac{n_{\rm e}}{T_{\rm e}}\frac{\partial \phi}{\partial \tau}+\nabla\cdot\left(n_{\rm e}\vec{u}_{\rm e}\right)=n_{\rm e}\nu_{\rm ion},
		\label{eq:hes1}
	\end{equation}
	\begin{equation}
		\frac{1}{\nu_{{\rm ela}}}
		\frac{\partial}{\partial \tau}\left(n_{\rm e}\vec{u}_{\rm e}\right)
		-n_{\rm e}\left[\mu \right]\nabla \phi+\left[\mu\right]\nabla\left(n_{\rm e}T_{\rm e}\right)=-n_{\rm e}\vec{u}_{\rm e},
		\label{eq:hes2}
	\end{equation}
	where $\tau$ is the pseudo time.
	Here, note that $n_{\rm e}$ is obtained from the ion density assuming a quasineutral plasma. 
	Taking $\phi$ and $n_{\rm e} \vec{u}_{\rm e}$ as the variables, flux Jacobian matrices can be formed and hence an approximate Riemann solver can be applied.
	Eqs. (\ref{eq:hes1}) and (\ref{eq:hes2}) are iterated until a steady state solution is obtained, which is achieved when the pseudo-time derivative terms become negligibly small values, say, to the level of round-off errors.
	By converting the elliptic equation in Eq. (5) into a hyperbolic system, the number of equations increases from one to three, so that the computational cost approximately triples if the number of iterations is unchanged.
	However, a preconditioning method is applied for the hyperbolic system approach to accelerate the iterative convergence.
	This preconditioning technique is designed to relax the stiffness arising from the magnetization of electrons \cite{Kawashima201559}.
	In the present simulation, the typical number of iterations required to converge the electron fluid calculation is $\sim10^4$, and the computational time for the electron fluid calculation in each physical time step is on the same order as that for the PIC calculations.
	Although the numerical method is different, the steady-state solution of the hyperbolic system is identical to the converged solution of the elliptic equation in Eq. (\ref{eq:elliptic}).
	
	\begin{table}
		\label{tab:condition}
		\caption{Thruster operation parameters assumed in the simulation.}
		\footnotesize
		\begin{center}
		\begin{tabular}{p{50mm}p{20mm}}
			\br
   	      Parameter & Value \\
			\mr
   	      Channel centerline diameter & 80 mm \\
      	   Channel width & 15 mm \\
            Mass flow rate  & 5.0 mg/s \\
	         Discharge voltage  & 300 V \\
         	Anode temperature & 850 K \\
			\br
		\end{tabular}
		\end{center}
	\end{table}
	\normalsize

\section{Numerical condition and method}

\subsection{Calculation condition}
	\label{sec:condition}
	The parameters for thruster configuration are presented in Table 1.
	The diameter of the channel centerline is 80 mm, which corresponds to the azimuthal length of 251 mm.
	Anode mass flow rate and discharge voltage are close to the operating condition of a SPT-100 thruster \cite{Manzella2001iepc}.
	Anode temperature is taken from Ref. \cite{ReidIEPC2015}.

\begin{figure}[t]
		\begin{center}
			\includegraphics[width=60mm]{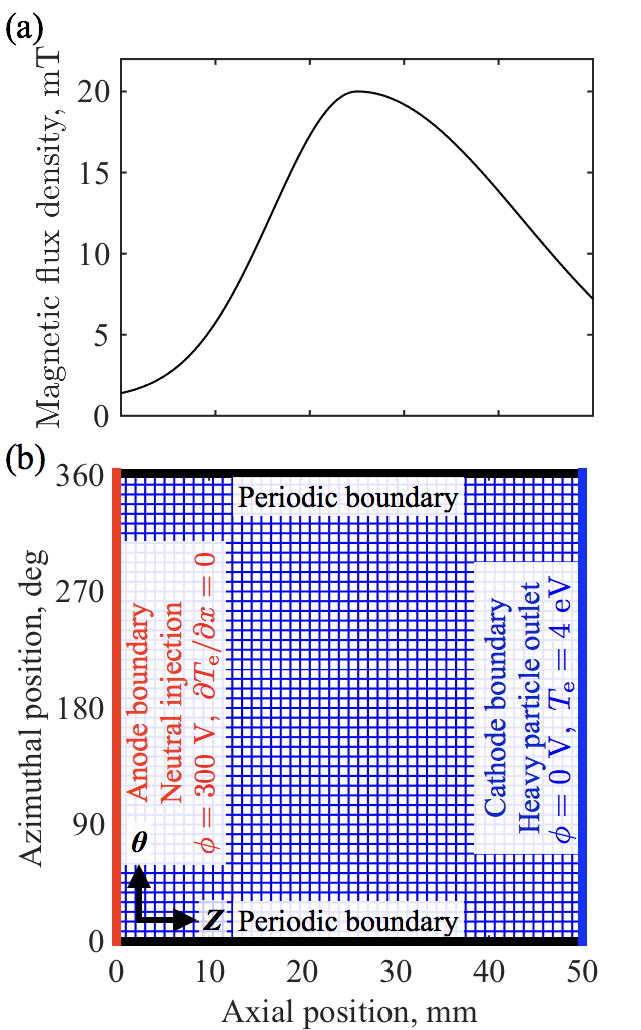}
		\end{center}
		\vspace{-2mm}
		\caption{(a) Axial distribution of magnetic flux density, which is assumed to be symmetric in the $\theta$-direction.
		(b) Calculation domain, boundary conditions, and computational mesh used in the axial-azimuthal simulation.}
		\label{fig:condition}
	\end{figure}

	The axial distribution of the radial magnetic flux density is shown in Fig. \ref{fig:condition}(a), while assuming that there are no fluctuations of the magnetic fields in the azimuthal direction.
	The magnetic flux density distribution is assumed by referring to the magnetic field configuration of a SPT-100 \cite{Mitrofanova}.
	The calculation domain, boundary conditions, and computational mesh are shown in Fig. \ref{fig:condition}(b).
	The calculation domain is set to cover the full cylinder of the channel centerline.
	In this simulation, a Cartesian coordinate is assumed, i.e., the finite curvature effects of the cylindrical coordinate are neglected, for simplicity.
	Neutral particles of xenon are injected from the anode boundary uniformly in the $\theta$-direction.
	Dirichlet conditions of space potential $\phi$ are imposed on the anode and cathode boundaries.
	The electron energy conduction to the anode is considered to be zero, and hence the gradient of electron temperature is set to zero at the anode boundary.
	The electron temperature at the cathode boundary is 4 eV.
	Periodic boundary conditions are used on the top and bottom boundaries for both heavy particles and electron fluid.
	The effects of the channel wall (in the radial direction) are ignored in the simulation for simplicity.
	The computational mesh of 48$\times$48 is used in this simulation.
	The simulation is initiated with an azimuthally uniform distribution, and transitions to a coherent structure within a few tenths of a millisecond.
	It was observed that the coherent structure is long-living after it forms.
	
	\begin{figure*}[t]
		\begin{center}
			\includegraphics[width=155mm]{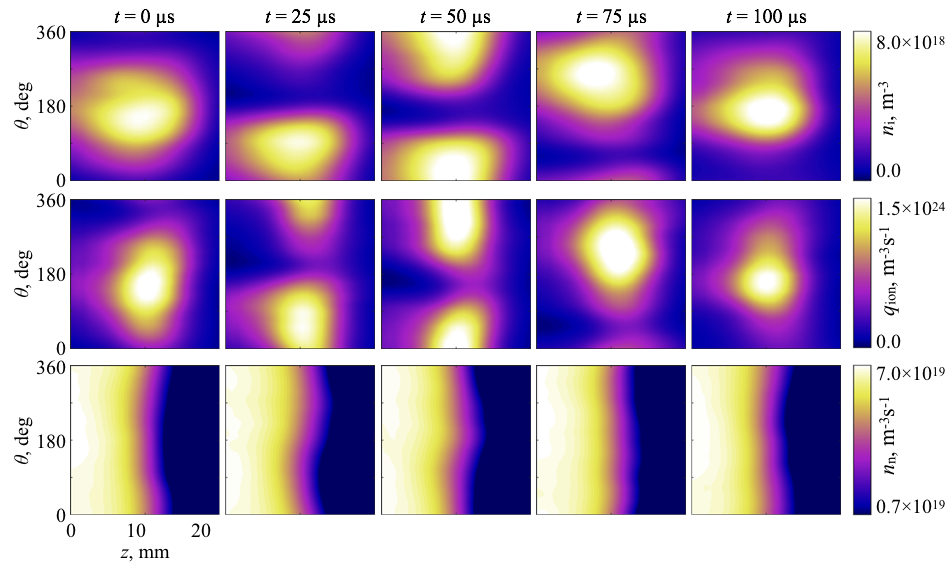}
		\end{center}
		\vspace{-3mm}
		\caption{Snapshots of (a) ion number density, (b) ionization rate, and (c) neutral number density. The contour in (c) is in logarithmic scale.
		In these distributions, the high-density regions are rotating in the $+E \times B$ direction.}
		\label{fig:move}
   \end{figure*}

\subsection{Numerical method}
	The ions and neutral atoms are modeled using a two-dimensional-three-velocity (2D3V) PIC method.
	The number of macroparticles is adjusted to make each cell contain $\sim200$ macroparticles on average regarding the cells in the domain, which means that approximately 4.6$\times$10$^5$ macroparticles are handled in the calculation domain for the 48$\times$48 grid.
	The time step for the PIC calculation is $\Delta t = 1.0 \times 10^{-8}$ s.
	Macroscopic quantities, such as the density and mean velocity, are obtained by weighting the particle information on the cell centers.
	
	For the electron fluid model with pseudo-time stepping, Steger-Warming's flux vector splitting \cite{Steger1981263} is employed as an approximate Riemann solver.
	A second-order spatial accuracy is attained by using a total variation diminishing monotonic upstream scheme for conservation laws (TVD-MUSCL) technique, coupled with a minmod limiter function.
	The lower-upper symmetric Gauss-Seidel (LU-SGS) method is used for the time integration with the Courant number of 2.0 \cite{YOON:1988aa}.
	The 1D electron energy conservation equation is calculated by applying second-order upwind method and 2nd-order central differencing.
	A fully implicit method using the direct matrix inversion is employed for stable computation.
	The time step of electron fluid calculation is set as $\Delta t_{\rm e} = 1.0 \times 10^{-9}$ s.
	This means that 10 iterations of electron fluid calculation are implemented in the single time step for PIC.
	Within this 10 iterations of electron's sub-loop, the HES in Eqs. (\ref{eq:hes1}) and (\ref{eq:hes2}) and the electron energy conservation in Eq. (\ref{eq:energy}) are calculated iteratively.
	In each iteration, the converged solutions for the $\phi$ and $\vec{u}_{\rm e}$ are computed in the HES.
	The obtained solutions of potential end electron flux are averaged in the azimuthal direction, and then used in the 1D electron energy conservation calculation.
	The electron mobility tensor is calculated only once at the first iteration of the electron's sub-loop.
	Note that it is found that the changes in the number of iterations of the electron's sub-loop do not vary the calculation results.
	
	\begin{figure*}[t]
		\begin{center}
			\includegraphics[width=150mm]{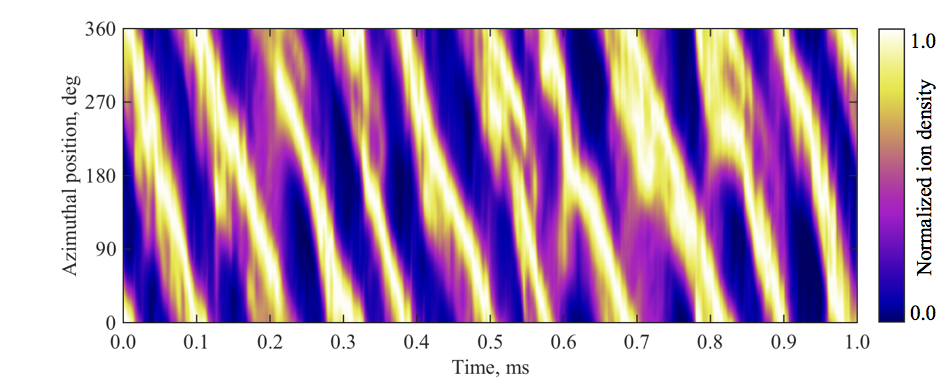}
		\end{center}
		\caption{Time history of azimuthal distribution of normalized ion number density.
		The cycle period and rotating speed are estimated as 100 $\mu$s and 2,500 m/s, respectively.}
		\label{fig:history}
   \end{figure*}

\section{Results}

\subsection{Rotating structure analysis}
	\label{sec:spoke}

	Snapshots of ion number density, ionization rate, and neutral number density distributions are shown in Fig. \ref{fig:move}.
	The results shown are zoomed-up in the ionization region that is in the upstream ($z<25$ mm) to illustrate the plasma structure propagating in the $\theta$ direction. 
	The spoke (here only one mode is observed) does not deform in the $z$ direction and coherently propagates in the E$\times$B direction.
	Note that the electron temperature was relatively static even with the time-dependent 1D electron energy solver, indicating that the breathing mode in the axial direction does not occur.
	The variation in the electron energy in the azimuthal direction is not required for the spoke to exist.
	It can be seen that the neutral atom front is curved due to the existence of non-uniformity of plasma density in the azimuthal direction, while $\partial_\theta n_{\rm n} = 0$ at the location where the plasma density is maximum. 
	Here, $\partial_\theta = \partial / \partial \theta$.
	The peaks in plasma density and ionization rate typically are in the same location because $\partial_\theta q_{\rm ion} \sim n_{\rm n} f_{\rm ion,Xe}(T_{\rm e}) \partial_\theta n_{\rm e}$. 
	In addition, it can be seen that the ionization rate elongates in the azimuthal direction depending on the neutral atom density profile both in $\pm E\times B$ directions, but the plasma structure preferentially and coherently propagates in E$\times$B direction. 
	This indicates that the spoke is not driven by the ionization instability {\em per se} and is driven preferentially in one direction by a different mechanism, namely, the electron dynamics.

	Figure \ref{fig:history} shows the spoke contour. 
	The plasma density is averaged in the axial direction to illustrate the time evolution of the plasma profile as a function of the azimuthal location. 
	The results shown are normalized for better tracking of the azimuthal motion of the wave.
	It can be seen from Fig. \ref{fig:history} that the plasma is coherently propagating in the same direction over a 10 cycles. 
	This result illustrates an unambiguous structure of a rotating spoke while there are some high-frequency signatures or small signals of different propagation speed.
	The spoke contour technique is described in Ref. \cite{SekerakTPS2015} to analyze rotating spokes in HETs.
	The phase velocity of the observed rotating structure is estimated by the propagation speed of the spoke contour.
	The rotating spoke frequency $\nu_{\rm rot}$ for the $m=1$ mode observed in our simulations is 10 kHz because the rotating structure circulates in the azimuthal direction 10 times over a period of 1.0 ms.
	In addition, the phase velocity is 2,500 m/s on average, because the azimuthal length of the computational domain is $L_\theta = 0.25$ m and $v_{\rm p} = L_\theta \nu_{\rm rot}$.

	\begin{figure}[t]
		\begin{center}
			\includegraphics[width=78mm]{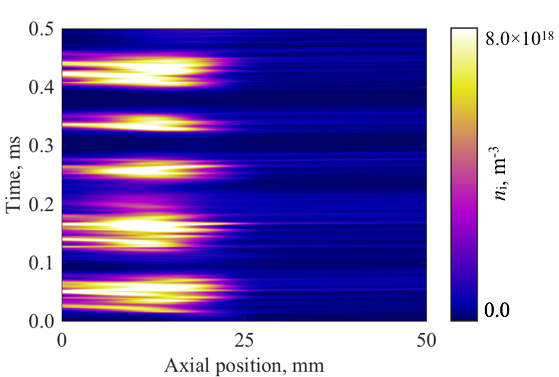}
		\end{center}
		\vspace{-2mm}
		\caption{Time history of axial ion number density distribution at $\theta$ of 180 deg.}
		\label{fig:z-t}
   \end{figure}
	
	The phase velocities of rotating spokes reported in the HET experiments are typically 1,500 -- 2,000 m/s in a 6 kW-level thruster \cite{SekerakTPS2015}, and 1,200 -- 2,800 m/s in a 100 W-level thruster \cite{EllisonPoP2012}.
	It can be seen that the phase velocity of the rotating structure obtained from our simulation agrees with experimental measurements of the rotating spokes.
	The phase velocity of the rotating spokes observed in magnetron discharges depend on the discharge condition and target material, and it is typically on the order of $10^3 - 10^5$ m/s \cite{Brenning:2013aa,Hecimovic:2016aa}.

	\begin{figure*}[t]
		\begin{center}
			\includegraphics[width=155mm]{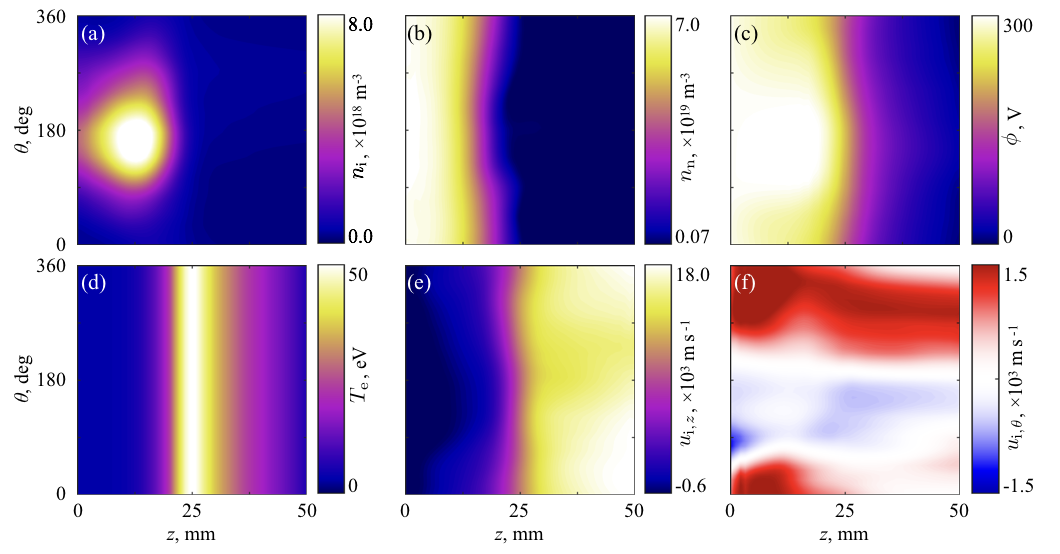}
		\end{center}
		\vspace{-3mm}
		\caption{Typical two-dimensional distributions of plasma properties. (a) ion number density, (b) neutral number density in logarithmic scale, (c) space potential, (d) electron temperature ($\theta$-symmetry assumed), (e) axial ion velocity, and (f) azimuthal ion velocity.
		The axial and azimuthal lengths of the domains are not to scale.
		}
		\label{fig:dist2d}
	\end{figure*}

	The observed rotating spoke shows the characteristics of an ionization oscillation because the ion density and the ionization rate are propagating simultaneously.
	However, the driving mechanism of the rotating spoke is unlikely to be ionization instability because the electron energy is uniform in the azimuthal direction and there are no oscillations in the electron energy.
	This was suggested by Hara et al. for their breathing mode studies \cite{Hara:2014aa}.
	To investigate the source of the rotating spokes, linear stability analysis is used to analyze the characteristics of the coherent structure in Sec. \ref{sec:linear}.
	
	Figure \ref{fig:z-t} shows the temporal evolution of the plasma density as a function of the axial position at a given azimuthal position. 
	It can be seen that the rotating spoke frequency is approximately 10 kHz and that plasma density profile stays similar over multiple cycles. 
	Hence, this also shows that the  rotating spoke is not growing nor being damped, namely, coherently propagating in the azimuthal direction. 
	It is to be noted that there is a slight time lag between the time when the ion density in the ionization region ($z<25$ mm) is peaked and that when the ion density in the acceleration region ($z>25$ mm) is peaked. 
	This is due to the ion velocity being small in the ionization region, and a finite time is required for the ions to be accelerated downstream in the $+z$ direction.

\subsection{Two-dimensional plasma properties}

	Figure \ref{fig:dist2d} shows instantaneous results of plasma properties at $t=100$ $\mu$s.
	As the rotating spoke propagates coherently, the snapshot is representative of what the spoke structure is.
	Additionally, the results capture the HET discharge plasma in that ionization region is in the upstream (near the anode) and the ions accelerate downstream (towards the cathode).
	The evolution of ion number density, neutral atom density, and ionization rate is shown in Fig. \ref{fig:move}.
	The peak value of the ion number density $n_{\rm i,p}$ is $\sim 8\times 10^{18}$ m$^{-3}$, which is almost five times larger than a typical $n_{\rm i,p}$ observed in axial-radial simulations \cite{Hofer:2007aa}.
	A large $n_{\rm i,p}$ in the present simulation is attributed to the non-uniformity in the azimuthal direction.
	The plasma density that is averaged in the azimuthal direction has a maximum of $4\times 10^{18}$ m$^{-3}$, which is closer to the previous simulation results.
	In addition, the plasma density may be overestimated because the radial ion flux toward the wall is not taken into account.
	The distribution of ionization rate $q_{\rm ion}$ is also non-uniform and enhanced in the high-density region of $n_{\rm i}$.
	The neutral number density $n_{\rm n}$ at the high-density region of $n_{\rm i}$ is slightly depleted compared with other azimuthal locations, due to the high ionization rate.
	
	The potential distribution is also nonuniform in the azimuthal direction (see Fig. \ref{fig:dist2d}(c)).
	A plateau region is observed in the potential distribution corresponding to the high-density region of ion number density.
	The peak electron temperature $T_{\rm e,p}$ is $\sim 50$ eV, which is almost twice of the one observed in the previous axial-radial simulation \cite{Hofer:2007aa}.
	The electron temperature in the present simulation may also be overestimated because the electron energy losses to the wall are not taken into account.
	
	As shown in Fig. \ref{fig:dist2d}(e), the axial ion mean velocity shows a typical HET-like profile. 
	Ions are accelerated downstream due to the potential drop while experiencing diffusion towards the anode as well. 
	The acceleration region (say, $z>25$ mm is relatively quasi-1D in that the ion acceleration occurs uniformly in the azimuthal direction. 
	Note that the anode sheath is not taken into account in the present calculations. 
	The reason why the axial ion mean velocity has some azimuthal variations is because the non-uniformity of the plasma potential structure. 
	For instance, at $\theta=180^\circ$ where the plasma density is largest, the potential drop between the plasma and the anode is also large. Hence, the ion diffusion occurs more towards the anode. 
	The effects of the anode sheath to the rotating spoke will be investigated in future work.
	Finally, the azimuthal ion mean velocity is shown in Fig. \ref{fig:dist2d}(f). 
	The ion mean velocity is slower than the propagation speed of the rotating spoke ($\sim 2,500$ m/s) where the plasma density is largest. 
	Hence, the rotating spoke is not due to ion convection in the azimuthal direction. 
	Another interesting observation is that the ions are moving in the $-E\times B$ ($+\theta$) direction. This is consistent with the unified spoke model suggested in Ref. \cite{Brenning:2013aa}, in which ions travel in the opposite direction of the electrons.
	At $z=50$ mm, the ion velocity in the azimuthal direction $u_{\rm i,\theta}$ is 1.5 km/s at most, which is much smaller than $u_{{\rm i},z}$ at the same axial position.
	Thus, despite the presence of azimuthally nonuniform distribution, ions are essentially accelerated in the axial direction.

\section{Driving mechanism of rotating spokes}
	\label{sec:linear}
	To investigate the driving mechanism of the rotating spoke, linear perturbation theories are used to obtain growth rates and phase velocity of the plasma wave using the results of the numerical simulations.
	Because the rotating spoke coherently propagates for a long time without being dissipated, it is likely that the rotating spoke experiences a linear instability that drives the structure, and the coherent wave propagation indicates nonlinear saturation of the linear growth.
	As the wave propagation will typically follow the phase velocity of the maximum growth rate, the linear instability analysis would be helpful in understanding what the source of rotating structure is.
	Moreover, the phase velocity predicted from theories can be directly compared to the simulation results to investigate the spoke mechanism.
	The procedures for applying linear analysis are as follows: 1) axial distributions of time-averaged plasma properties are obtained from the numerical simulation, 2) the axial profile of the plasma properties are inserted into various dispersion relations that are evaluated locally, and 3) (spatially) local growth rate and real frequency, thus, the phase velocity of the dispersion relation are assessed.
	Here the phase velocity in the azimuthal direction of $m=1$ is compared between the numerical analysis and linear theories.

\subsection{Linear theories for rotating spoke analysis}
	Several dispersion relations are considered to be related to the azimuthally rotating spokes.
	The characteristics of rotating spokes due to closed drift in HETs have been investigated since the 1970s.
	Morozov et al. first implemented a linear analysis in the azimuthal direction in a HET based on a model of quasineutral, two-fluid, collisionless plasma without electron inertia.
	This model identifies the existence of azimuthal fluctuations driven by the combination of magnetic field and density gradients, and the criterion to cause this instability is simply expressed as 
	\begin{equation}
		\frac{B_r}{n_{\rm e}}\frac{\partial}{\partial x}\left(\frac{n_{\rm e}}{B_r}\right)>0.
		\label{eq:Morozov}
	\end{equation}
	For typical HETs, an instability related to this criterion is induced where $\partial B_r/\partial x<0$ \cite{ChoueiriPoP2001}.
	Esipchuk and Tilinin further developed this theory and derived a dispersion relation of a drift instability resulting from the gradients of plasma density and magnetic flux density in the axial direction \cite{Esipchuk:1976}.
	Esipchuk's dispersion relation is described as follows:
	\begin{eqnarray}
		\omega=k_{z}u_{{\rm i},z}
		-\frac{k_\perp^2 u_{{\rm i},z}^2}{2k_\theta \left(u_{{\rm e},\perp}-u_B\right)}
		\pm\frac{k_\perp u_{{\rm i},z}^2}{2 \left(u_{{\rm e},\perp}-u_B\right)}
		\hspace{10mm}\nonumber\\
		\times \sqrt{1+\left(\frac{k_z}{k_\theta}\right)^2
		-4\frac{u_{{\rm e},\perp}-u_B}{u_{{\rm i},z}}\left(\frac{k_z}{k_\theta}-\frac{u_{{\rm e},\perp}}{u_{{\rm i},z}}\right)},
		\label{eq:Esipchuk}
	\end{eqnarray}
	where $k_\perp=\sqrt{k_\theta^2+k_z^2}$, and $u_B$ and $u_{\rm e,\perp}$ are defined as follows:
	\begin{equation}
		u_B=\frac{u_{{\rm i},z}^2}{\omega_{\rm c,i}B}\frac{\partial B}{\partial z},
	\end{equation}
	\begin{equation}
		u_{\rm e,\perp}=\mu_\perp\frac{\partial \phi}{\partial z}-\frac{\mu_\perp}{n_{\rm e}}\frac{\partial}{\partial z}\left(n_{\rm e}T_{\rm e}\right).
	\end{equation}
	Here $\omega_{\rm c,i}$ is the angular frequency of ion-cyclotron motion.

	More recently, Frias et al. have proposed a perturbation theory which is based on the drift instability that accounts for the compressibility of electrons \cite{FriasPoP2012}.
	This gradient drift wave theory was further extended to include the electron temperature perturbation.
	We focus on the linear dispersion relation without electron temperature fluctuation in Ref. \cite{FriasPoP2012} as it was observed that the result will not differ too drastically by including the electron temperature contributions.
	Frias's gradient drift theory is written as
	\begin{eqnarray}
		\omega=k_{z}u_{{\rm i},z}
		+\frac{k_\perp^2 u_{\rm s}^2}{2\left(\omega_*-\omega_D\right)}
		\pm\frac{k_\perp^2 u_{\rm s}^2}{2\left(\omega_*-\omega_D\right)}
		\nonumber\\
		\times \sqrt{1
		+4\frac{k_z u_{{\rm i},z}}{k_\perp^2u_{\rm s}^2}\left(\omega_*-\omega_D\right)
		-4\frac{k_\theta^2}{k_\perp^2}\rho_{\rm s}^2\Delta}.
		\label{eq:Frias}
	\end{eqnarray}
	Here $u_{\rm s}=\sqrt{eT_{\rm e}/m_{\rm i}}$ is the ion sound speed, and $\omega_{*}$, $\omega_{\rm D}$, $\rho_{\rm s}$, and $\Delta$ are respectively defined as follows:
	\begin{equation}
		\omega_{*}=-k_\theta\frac{T_{\rm e}}{Bn_{\rm e}}\frac{\partial n_{\rm e}}{\partial z},
	\end{equation}
	\begin{equation}
		\omega_{\rm D}=-2k_\theta\frac{T_{\rm e}}{B^2}\frac{\partial B}{\partial z},
	\end{equation}
	\begin{equation}
		\rho_{\rm s}^2= \frac{m_{\rm i}T_{\rm e}}{eB^2},
	\end{equation}
	\begin{equation}
		\Delta = \left(\frac{2}{B}\frac{\partial B}{\partial z}-\frac{1}{T_{\rm e}}\frac{\partial \phi}{\partial z}\right)
		\left(\frac{1}{n_{\rm e}}\frac{\partial n_{\rm e}}{\partial z}-\frac{2}{B}\frac{\partial B}{\partial z}\right).
	\end{equation}
	Note that for both Eqs. (\ref{eq:Esipchuk}) and (\ref{eq:Frias}) the growth rate appears when the argument inside the square root is negative. 
	If it is positive, then the real frequency will have two roots without any growth or damping. 
	However, the wave with a positive growth likely dominates the entire plasma wave structure. 
	Hence, the focus must be on whether a positive growth rate exists and the phase velocity where the growth rate is positive.

   \begin{figure}[t]
   	\begin{center}
   		\includegraphics[width=70mm]{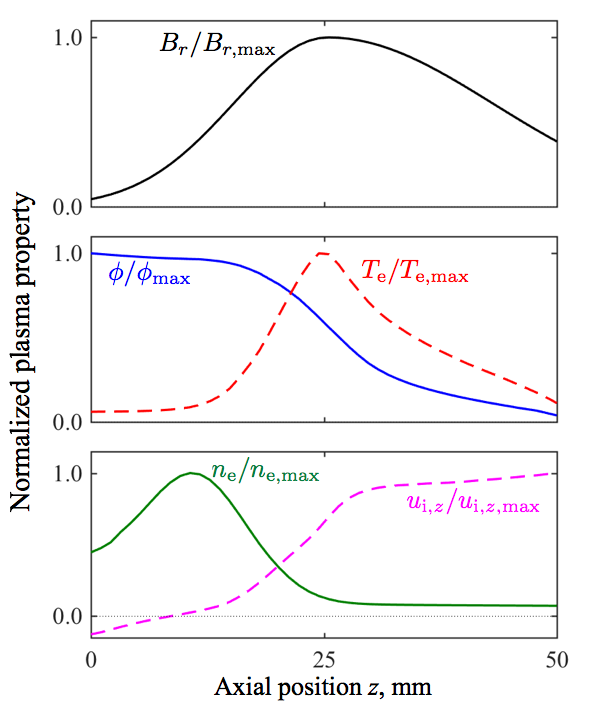}
	   \end{center}
	   \vspace{-2mm}
   	\caption{Axial distributions of normalized plasma properties.
	The distributions are averaged in the azimuthal direction and further averaged within $0.5\ {\rm ms}\leq t\leq 1.0\ {\rm ms}$.
	The maximum values of the properties are as follows: $B_{r,{\rm max}}=20$ mT, $\phi_{\rm max}=299$ V, $T_{\rm e,max}=50.3$ eV, $n_{\rm e,max}=4.10\times 10^{18}$ m$^{-3}$, $u_{{\rm i},z,{\rm max}}=2.09\times 10^4$ m/s.}
   	\label{fig:dist1d}
   \end{figure}

	Following Sekerak's work \cite{SekerakThesis}, we also compare the two phase velocities predicted by the CIV model and electrostatic ion cyclotron oscillations (ESIC), which are given by,
	\begin{equation}
		v_{\rm CIV} = \sqrt{\frac{2e\varepsilon_{\rm ion}}{m_{\rm i}}},
		\label{eq:CIV}
	\end{equation}
	and
	\begin{equation}
		\omega^2=k_\theta^2u_{\rm s}^2+\omega_{\rm c,i}^2,
		\label{eq:ESIC}
	\end{equation}
	respectively.
	The dispersion relations derived above are applied to the numerical simulation results.
	By inserting the local plasma properties into the dispersion relations in Eqs. (\ref{eq:Esipchuk}) and (\ref{eq:Frias}), the growth rate is calculated from the imaginary part, and the phase velocity is obtained from the real part of the frequency.
	Eqs. (\ref{eq:CIV}) and (\ref{eq:ESIC}) do not have the imaginary parts, and hence only phase velocity is obtained.
	The required plasma properties in these equations are magnetic flux density $B$, space potential $\phi$, electron temperature $T_{\rm e}$, electron number density $n_{\rm e}$, and axial ion velocity $u_{{\rm i}, z}$.
	The two-dimensional plasma properties are averaged in time and in the azimuthal direction, generating a steady-state axial profile of the plasma parameters, which are shown in Fig. \ref{fig:dist1d}.
	Concerning the wave number, in the azimuthal direction, $k_\theta = 1$ is assumed to be consistent with the numerical simulation.
	In the axial direction, $k_z = 0.1k_\theta$ is assumed referring to the analysis in Ref. \cite{ChoueiriPoP2001}. 
	$k_\perp$ is calculated by $k_\perp=\sqrt{k_\theta^2+k_z^2}$.

   \begin{figure}[t]
   	\begin{center}
   		\includegraphics[width=75mm]{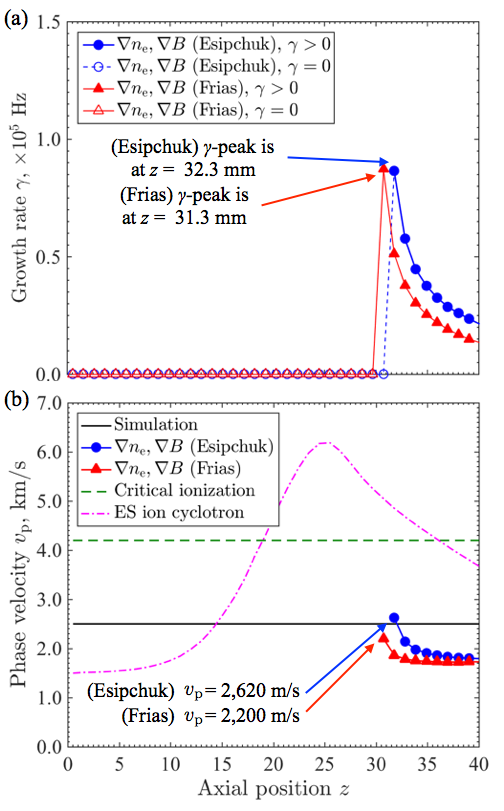}
	   \end{center}
	   \vspace{-2mm}
   	\caption{Axial distributions of growth rates and phase velocities derived by various dispersion relations.
			(a) Growth rates obtained by the Esipchuk's and Frias's dispersion relations.
			(b) Phase velocities obtained by the simulation, Esipchuk's and Frias's dispersion relations, CIV, and electrostatic ion cyclotron velocity.
			Phase velocity for $\gamma=0$ is not shown in Esipchuk's and Frias's dispersion relations because the wave with the maximum growth rate will be dominant in the system.}
   	\label{fig:dispersion}
   \end{figure}

\subsection{Linear-stability analysis results}
	By substituting the axial distributions of plasma properties into each dispersion relation, the growth rate and phase velocity are analyzed.
	The growth rate obtained from the Esipchuk's equation in Eq. (\ref{eq:Esipchuk}) and Frias's equation in Eq. (\ref{eq:Frias}) are shown in Fig. \ref{fig:dispersion}(a).
	In both results of Esipchuk's and Frias's equations, large growth rates are observed in the downstream region of $z>25$ mm.
	Particularly, in Eq. (\ref{eq:Frias}), the growth rate becomes largest when $\omega_* = \omega_D$, meaning that the gradients of plasma density and magnetic field strengths play an important role in the gradient drift instability in the azimuthal direction.
	As shown in Fig. \ref{fig:dist1d}, $\omega^* \sim 0$ at $z=12$ mm, and $\omega^*$ has a maximum at $z>12$ mm, and $\omega^* \rightarrow 0$ as $z \rightarrow \infty$, while $\omega_D = 0$ at $z=25$ mm where $B$ is at maximum and $\omega_D$ increases at $z>25$ mm. 
	Hence, the point of $\omega^* = \omega_D$ exists at $z>25$ mm, namely, downstream of the maximum magnetic field.
	The trend of large growth rate in the plume region is consistent with the results of Choueiri's linear analysis of SPT-100 thruster\cite{ChoueiriPoP2001}, and Frias's linear analysis of various thrusters\cite{FriasPoP2013}.
	In the present analysis, the growth rate calculated using Esipchuk's and Frias's equations show the peaks at $z=32.3$ mm and $z=31.3$ mm, respectively.
	Based on Esipchuk's and Frias's dispersion relations, it can be considered that these locations are the source of an azimuthally rotating oscillations.

	\begin{table}
		\label{tab:phase}
		\caption{Phase velocities predicted by simulation and various dispersion relations.}
		\footnotesize
		\begin{center}
		\begin{tabular}{p{15mm}p{15mm}p{15mm}p{20mm}}
			\br
		    & $z$, mm & $v_{\rm p}$, m/s  &  Difference from Simulation      \\
			\mr
			Simulation  & -- & 2,500 & -- \\
			Esipchuk & 32.3    & 2,620 & 4.8\%    \\
			Frias  & 31.3  & 2,200 & 12.0\%   \\
			CIV  & --  & 4,200 & 68.0\%   \\
			ESIC  & 12.0  & 1,840 & 26.4\%   \\
			\br
		\end{tabular}
		\end{center}
	\end{table}
	\normalsize

	The phase velocities calculated from the simulation, Esipchuk's and Frias's dispersion relations, CIV, and ESIC are shown in Fig. \ref{fig:dispersion}(b).
	The phase velocity of the rotating structure in the simulation is 2,500 m/s.
	Using Esipchuk's dispersion relation, the phase velocity is 2,620 m/s where the growth rate is largest at $z=32.3$ mm.
	Similarly, Frias's dispersion relation results in a phase velocity of 2,200 m/s at $z=31.3$ mm.
	Here, we consider that the azimuthal structure propagates with the phase velocity where the growth rate is largest.
	CIV is a specific value for the gas species and is 4,200 m/s for xenon.
	Lastly, concerning the ESIC, the phase velocity is 1,840 m/s if it refers the value at $z=12.0$ mm where the ion number density shows the peak.
	ESIC model also gives a good agreement with the numerical simulation, however, the hybrid-PIC model assumes that the ions are non-magnetized.
	Hence, it is unlikely that ESIC is the driving mechanism of rotating spokes.
	
	The phase velocities predicted by the simulation and various dispersion relations are compared in Table 2.
	The phase velocity of Esipchuk's dispersion relation shows the best match with the simulation with the difference of 4.8\%.
	Frias's dispersion relation also showed the phase velocity close to the simulation with the difference of 12\%.
	These dispersion relations are both based on the drift instability owing to the gradients of plasma density and magnetic flux density.
	Therefore, it is considered that the drift instability is the driving mechanism of the rotating spoke observed in the numerical analysis.
	In Sec. \ref{sec:spoke}, it was shown that rotating spoke is associated with ionization oscillations.
	However, linear analysis suggests that ionization instability is not the cause of the rotating spoke.
	The results of this research supports the hypothesis that the rotating spoke is induced by the drift instability that occurs downstream of the thruster.
	
   \begin{figure}[t]
   	\begin{center}
   		\includegraphics[width=80mm]{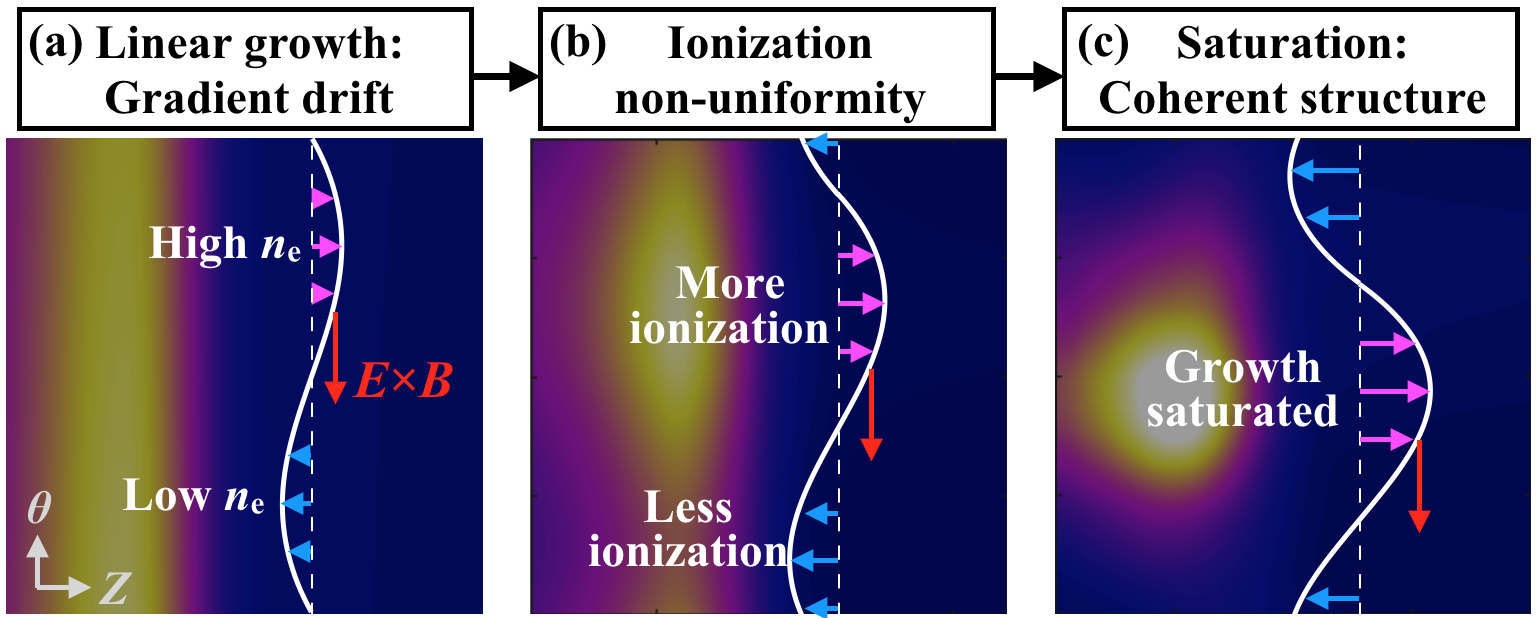}
	   \end{center}
	   \vspace{-2mm}
   	\caption{Hypothesized multidimensional spoke model.
			The white line in each figure represents azimuthal electron density distribution.
			(a) Starting from the azimuthally uniform state, a small oscillation in the electron density is induced by the drift instability at downstream.
			(b) The oscillation grows by non-uniform ionization.
			(c) The oscillation growth is saturated and a coherent wave propagates in the azimuthal direction with a steady phase speed.}
   	\label{fig:mechanism}
   \end{figure}

\subsection{Multidimensional spoke model}
	A multidimensional spoke model is shown in Fig. \ref{fig:mechanism}.
	Starting from azimuthally uniform state, a small plasma wave propagates in the $E_z\times B_r$ direction and grows linearly due to the gradient drift instability at downstream.
	The linear growth of the gradient drift wave will result in fluctuations of plasma density and space potential.
	The azimuthal variation in plasma density causes non-uniformities in ionization rate and neutral atom density, which in turn enhances the non-uniformity of plasma density.
	The non-uniform ionization profile penetrates in the axial direction.
	After a certain duration, the azimuthal non-uniformity reaches the anode boundary.
	At this time, non-uniformities in the space potential and neutral atom density no longer grow, owing to the uniformly fixed potential and neutral atom flow rate at the anode.
	Finally, the instability saturates and the non-uniformity of the plasma constituents becomes fully developed.
	A coherent structure of azimuthal wave is formed and propagates in the azimuthal direction with a constant phase velocity.
	After the gradient drift instability nonlinearly saturates, the propagation velocity of the coherent structure is close to the phase velocity of the gradient drift instability.
	This indicates that the azimuthal transport of the coherent wave is driven by the gradient drift wave, whereas the non-uniform ionization plays an important role in the  wave amplification and information travel in the axial direction.
	This hypothesized multidimensional spoke model is consistent with the results of the numerical simulation and linear theories.
	A detailed numerical simulation on the evolution process may validate this model. 
	To observe a smooth transition from an azimuthally uniform state to a rotating spoke state, a careful choice of the initial condition is required. 
	In addition, the effects of the empirical anomalous electron mobility profile on rotating spokes as well as the induced electron transport due to rotating spokes will be reserved for future work.

\section{Conclusion}
	An axial-azimuthal two-dimensional simulation of a HET is performed using a particle-fluid hybrid model in which the electron fluid model is solved with a hyperbolic system approach.
	In the absence of axial ionization oscillations, a low-frequency coherent structure that propagates in the azimuthal direction is observed.
	It was confirmed that the azimuthally rotating spokes propagate coherently for a long time exceeding 1 ms without any dissipation or dispersion of the plasma wave.
	The electron fluid model using drift-diffusion model yields a large anisotropic diffusion that is numerically difficult to solve. 
	Mathematically, the second-order elliptic partial differential equation for the plasma potential becomes ill-conditioned and conventional iterative methods suffer from numerical convergence. 
	The hyperbolic system approach introduces a pseudo-time stepping term so that the elliptic equation is converted into a hyperbolic equation system. 
	The steady-state solution of the hyperbolic approach is identical to the converged solution of the elliptic equation.
	The hyperbolic system approach is a robust computational method for magnetized electron fluid calculation, where the issues of numerical instability and stiffness are avoided by using several numerical techniques.
	
	In our calculations of a SPT-100 thruster, a coherent structure of $m=1$ wave is observed in the azimuthal direction.
	The propagation speed was 2,500 m/s. 
	In the presence of plasma density peak in the azimuthal direction, the neutral atom density also shows non-uniformity. 
	This is a feature of the ionization oscillation. 
	However, the electron temperature is assumed to be uniform in the azimuthal direction in our calculations. 
	Thus, the rotating spoke is not induced by ionization instabilities.
	The propagation speed of the rotating spoke obtained from the numerical simulation agrees with experimental observations.
	
	To investigate the driving mechanism of the rotating spoke, a comparison between the numerical analysis and several linear stability analyses is conducted.
	Axial distributions of steady-state plasma properties are obtained from the numerical simulation and are used as input in the linear analyses.
	The dispersion relations based on the drift instability show growth rate of azimuthal oscillation wave in the plume region downstream of the peak of the radial magnetic field.
	The phase velocity predicted by the linear analysis of recent drift instability model is 2,200 m/s.
	The close match of phase velocities between the numerical simulation and linear analysis suggest that the gradient drift instability is a potential mechanism of the rotating spoke.
	Finally, a multidimensional spoke model is proposed based on the results of numerical simulation and linear theory of gradient drift instability.
	The hypothesized model shows that the azimuthal transport of the coherent wave is driven by the gradient drift wave, whereas the non-uniform ionization plays an important role in the growth enhancement and information travel in the axial direction.

\section*{Acknowledgments}
	This work was supported in part by JSPS KAKENHI Grant Number JP17K14873.
 
\section*{References}
\bibliographystyle{iopart-num}
\bibliography{reference}

\end{document}